\documentstyle[12pt]{article}

\begin{document}

\def\thefootnote{\fnsymbol{footnote}}
\baselineskip 7.5 mm

\begin{flushright}
\begin{tabular}{l}
ITP-SB-94-30    \\
hep-ph/9406367 \\
February, 1994
\end{tabular}
\end{flushright}

\vspace{8mm}

\setcounter{footnote}{0}

\begin{center}
{\Large \bf Some model-independent properties of quark mixing}\footnote{
talk presented at the {\it Yukawa couplings and the origin of mass}
workshop, February 1994, University of Florida, Gainesville, Florida} \\

\vspace{12mm}

\setcounter{footnote}{6}

Alexander Kusenko   \\
Institute for Theoretical Physics \\
State University of New York at Stony Brook \\
Stony Brook, NY 11794-3840\footnote{address after September 1, 1994:
Department of Physics, University of Pennsylvania, Philadelphia, PA 19104} \\

\vspace{24mm}

{\bf Abstract }

\end{center}

We discuss some new invariants of quark mixing and show their
usefulness with a simple example.  We also present some other new tools for
analyzing quark mixing.

\vfill
\pagestyle{empty}
\newpage

\pagestyle{plain}
\pagenumbering{arabic}
\renewcommand{\thefootnote}{\arabic{footnote}}
\setcounter{footnote}{0}

\section{Invariants of quark mixing and their applications}

In recent years the increasingly accurate data on quark mixing has stimulated
interest in possible predictions concerning quark mass matrices.
The parameters of the Cabibbo-Kobayashi-Maskawa mixing matrix $V$
are now known
reasonably well \cite{3_data}.  This determination has been made possible
partly by the finding that there are only three generations of usual
standard-model
fermions (with corresponding light or massless neutrinos).  Since the
diagonalization of the quark matrices in the up and down sectors determines
$V$, one can work back from the knowledge of $V$ to put constraints on the
possible forms of (original, nondiagonal) quark mass matrices.  However,
the data on quark mixing determines these mass matrices
only up to an arbitrary unitary similarity transformation.  This is a result of
the fact that if the up and down
quark mass matrices, $M_{u}$ and $M_{d}$, are both acted on by the same
unitary operator $U_{0}$ according to

\begin{equation}
M_{u,d} \ \rightarrow \ U_{0} \: M_{u,d} \: U_{0}^{\dag}  \label{u0}
\end{equation}
then the mixing matrix $V$ remains unchanged.
There have been many attempts to study specific assumed forms for quark
matrices.  While this is worthwhile, it is desirable to
express the constraints from data on $V$  on the quark mass matrices in an
invariant form.  In Refs. \cite{3_Jarlskog,3_Branco}  certain invariant
functions of the quark mass matrices
$I_{pq}$ were introduced, which are expressed in terms of the quark masses
squared and the $|V_{ij}|$:
\begin{equation}
I_{pq} = Tr(H_{u}^{p} \: H_{d}^{q}) =\sum_{ij} \: (m^{(u)}_{i})^{2 p}
(m^{(d)}_{j})^{2 q} \: |V_{ij}|^{2}
\label{ieq}
\end{equation}
where $H_{q}=M_{q} M_{q}^{\dag}$, and $m_{i}^{(u)}$ and $m_{i}^{(u)}$ are
the masses of quarks in the ``up'' and ``down'' charge sectors.

In this paper we introduce some new invariants of the quark mass
matrices (with respect to the transformation (\ref{u0})) which can be
expressed in terms of the measurable quantities only.
These new invariants help one simplify the algebraic expressions
which relate the elements of the quark mass matrices to the data.  This
important advantage allows for some new uses of the invariants which we
will illustrate with some examples.

We introduce the following new invariants of the transformation (\ref{u0}):

\begin{equation}
K_{pq}(\alpha,\beta)= det ( \alpha H_{u}^{p}+ \beta H_{d}^{q} ) \label{K}
\end{equation}
where $ p,q,\alpha, \beta  \ \neq 0 $.

 The hermitian matrices $H_{u}$
and $H_{d}$ can be diagonalized by a unitary similarity transformation:
\begin{eqnarray}
\left \{ \begin{array}{l}
         U_{u} H_{u} U_{u}^{\dag} = D_{u}  \\
         U_{d} H_{d} U_{d}^{\dag} = D_{d}
         \end{array} \right.
\end{eqnarray}
where $D_{q}=diag((m_{1}^{(q)})^2,(m_{2}^{(q)})^2,(m_{3}^{(q)})^2)$ are the
diagonal matrices of the quark masses squared.

The mixing matrix $V$ can be written then as:
\begin{equation}
V= U_{u} U_{d}^{\dag}
\end{equation}

In order to find an expression for $K_{pq}$ in terms of $U_{ij}$ we will need
the following

\vspace{4mm}

\noindent {\bf Theorem}

{\em If $A$ and $B$ are two $ 3\times 3$ matrices
such that $ det(A) \neq 0$ and $ det(B) \neq 0$ then the following relation
holds:

\begin{equation}
det(A+B)= det(A) + det(B) + det(A) \: Tr(A^{-1} B)+ det(B) \: Tr(A B^{-1})
\label{theorem}
\end{equation}}

\vspace{4mm}

\noindent {\em Proof:}

We denote the elements of matrices $A$ and $B$ by $A_{ij}$ and
$B_{ij}$  correspondingly.  Their co-factors (which are equal to the
corresponding
minors, up to sign) will be written as $\hat{A}_{ij}$ and $\hat{B}_{ij}$.
Then each determinant may be decomposed in a sum (Laplace expansion):

\[
det(A)= \sum_{i} A_{ij} \hat{A}_{ij}=\sum_{j} A_{ij} \hat{A}_{ij}  \]
\[ det(B)= \sum_{i} B_{ij} \hat{B}_{ij}=\sum_{j} B_{ij} \hat{B}_{ij}
\]

By definition, the determinant of a $3\times 3$ matrix is a sum of $3!=6$
terms:

\begin{equation}
det(A+B)= \sum (-1)^{r} (A_{1 k_{1}} + B_{1 k_{1}} )
(A_{2 k_{2}}+ B_{2 k_{2}} ) (A_{3 k_{3}} + B_{3 k_{3}})  \label{det}
\end{equation}
where $r$ is the sign of the permutation $(^{1}_{k_{1}} \
^{2}_{k_{2}} \ ^{3}_{k_{3}})$.

The terms in the sum (\ref{det}) which contain only the elements of $A$
can be arranged as $det(A)$.  Similarly, the terms containing only
$B$'s give $det(B)$.  The terms containing one element of $A$ multiplied by
two elements of $B$, or visa versa, can be rewritten as:

\begin{equation}
\sum_{i,j} (A_{ij} \hat{B}_{ij}+B_{ij} \hat{A}_{ij})  \label{ab}
\end{equation}

We can now use an identity:

\[ (A^{-1})_{ij}= \frac{1}{det(A)} \hat{A}_{ji}  \]
to rewrite (\ref{ab}) as:

\[
 \sum_{i,j} (A_{ij} \hat{B}_{ij}+B_{ij} \hat{A}_{ij})
= det(B) \: \sum_{ij} A_{ij} (B^{-1})_{ji}+
det(A) \: \sum_{ij} (A^{-1})_{ji} B_{ij}=   \]

\[ det(A) \: Tr(A^{-1} B)+
det(B) \: Tr(A B^{-1})  \]

Altogether we get

\[ det(A+B)= det(A) + det(B) + det(A) \: Tr(A^{-1} B)+ det(B) \:
Tr(A B^{-1}) \]
which is the statement of the theorem (\ref{theorem}).  This completes the
proof.

  Theorem (\ref{theorem}) may be easily generalized to the case of
$2\times 2$ matrices, in which case the last two terms in (\ref{theorem})
are equal and correspond
to a redundant counting of the same terms in a sum similar to (\ref{det}).
Thus for the $2\times 2$ matrices we get:

\begin{equation}
\begin{array}{c}
det(A+B)= det(A) + det(B) + det(A) \: Tr(A^{-1} B) \equiv  \\
 det(A) + det(B) + det(B) \: Tr(A B^{-1})
\end{array}
\end{equation}

The immediate consequence of equations (\ref{theorem}) and
(\ref{ieq}) is the following relation:

\begin{eqnarray}
K_{pq}(\alpha, \beta) \ \equiv \ det(\alpha H_{u}^{p}+\beta H_{d}^{q}) \ =
\ \ \ \ \ \ \ \ \ \ \ \ \ \ \ \ \ \ \ \ \ \ \ \ \ \    \nonumber \\
\alpha^{3} \ (m_{1}^{(u)} m_{2}^{(u)} m_{3}^{(u)})^{2p} \
[\: 1+ (\beta/\alpha) \sum_{ij} \: [(m_{j}^{(d)})^{2q}/(m_{i}^{(u)})^{2p}]\:
U_{ij} \: ]   +  \label{kdm} \\
\beta^{3} \ (m_{1}^{(d)} m_{2}^{(d)} m_{3}^{(d)})^{2q} \
[\: 1+  (\alpha/\beta) \sum_{ij} \: [(m_{j}^{(u)})^{2p}/(m_{i}^{(d)})^{2q}] \:
U_{ij} \: ]   \nonumber
\end{eqnarray}

We also notice that

\begin{eqnarray}
K_{pq}(\alpha, \pm \beta) =
\alpha^3 (m_{1}^{(u)} m_{2}^{(u)} m_{3}^{(u)})^{2p} (1 \pm (\beta/\alpha)
I_{(-p) \, q}) \nonumber \\
\pm
\beta^3 (m_{1}^{(d)} m_{2}^{(d)} m_{3}^{(d)})^{2q} (1 \pm (\alpha/\beta)
I_{p \, (-q)}) \label{ij}
\end{eqnarray}

Any four independent invariants from the set $\{I_{pq}, K_{pq} \}$ contain all
the physical information about the CKM matrix.

If the mass matrices are assumed to be hermitian, one can introduce a similar
set of invariants:

\begin{equation}
\begin{array}{l}
\tilde{I}_{pq}=Tr ( \: M_{u}^{p} \:  M^{q}_{d} \:  )  \\
\tilde{K}_{pq}(\alpha,\beta)= det ( \alpha M_{u}^{p}+ \beta M_{d}^{q} )
\end{array}
\label{tildes}
\end{equation}

The formulae, similar to (\ref{ieq}), (\ref{kdm}) and (\ref{ij}),
will also hold for $\tilde{I}, \tilde{K}$:

\begin{equation}
\tilde{I}_{pq} = Tr(M_{u}^{p} \: M_{d}^{q}) =\sum_{ij} \: (m^{(u)}_{i})^{p}
(m^{(d)}_{j})^{q} \: |V_{ij}|^{2}
\label{tilde_ieq}
\end{equation}

\begin{eqnarray}
\tilde{K}_{pq}(\alpha, \beta) \ \equiv \ det(\alpha M_{u}^{p}+\beta M_{d}^{q})
\ =
\ \ \ \ \ \ \ \ \ \ \ \ \ \ \ \ \ \ \ \ \ \ \ \ \ \    \nonumber \\
\alpha^{3} \ (m_{1}^{(u)} m_{2}^{(u)} m_{3}^{(u)})^{p} \
[\: 1+ (\beta/\alpha) \sum_{ij} \: [(m_{j}^{(d)})^{q}/(m_{i}^{(u)})^{p}]\:
U_{ij} \: ]   +  \label{tilde_kdm} \\
\beta^{3} \ (m_{1}^{(d)} m_{2}^{(d)} m_{3}^{(d)})^{q} \
[\: 1+  (\alpha/\beta) \sum_{ij} \: [(m_{j}^{(u)})^{p}/(m_{i}^{(d)})^{q}] \:
U_{ij} \: ]   \nonumber
\end{eqnarray}

The odd powers of the mass eigenvalues may appear in some
of the $\tilde{I}$ and $\tilde{K}$-type invariants.  As usual, the signs of the
fermion masses are ambiguous.  However, for a given model, different choices of
signs will in general result in different predictions for the CKM matrix.  This
is because in general the mass matrices do not commute with all of the diagonal
matrices
of the form $diag(\pm 1, \pm 1, \pm 1)$, where the $+$ and $-$ signs are chosen
arbitrarily but so as to not get the plus or minus identity.
We discussed the effect of the sign choices on the CKM mixing parameters
on the example of some particular
model \cite{3_ks_model}.  Once the choice of signs for the
fermion eigenvalues (determined in practice by the best fit to the data) is
made, there is no further sign-related ambiguity in the $\tilde{I}$ and
$\tilde{K}$-type invariants.

Now we would like to illustrate the usefulness of the invariants discussed
above.  As an example, we will take the model proposed in \cite{3_ks_model}.
This model gives predictions for the low energy data on fermion masses and
mixing which are in reasonable agreement with experiment.  This model is
formulated in the context of an SO(10) supersymmetric grand unified theory.
The model has the following Yukawa matrices at the GUT scale:

\begin{equation}
M_u=  \left (\begin{array}{ccc}
                  0 & A_u & 0 \\
                  A_u & B_u & 0 \\
                  0 & 0 & C_u \end{array}   \right  )
\label{yu}
\end{equation}

\vspace{4mm}

\begin{equation}
M_d=     \left ( \begin{array}{ccc}
                  0 &  A_d e^{i\phi} &  0 \\
                  A_d e^{-i\phi} & B_d e^{i\theta} & B_d \\
                  0 & B_d & C_d \end{array}  \right )
\label{yd}
\end{equation}

\begin{equation}
M_e=     \left ( \begin{array}{ccc}
                  0 &  A_d e^{i\phi} &  0 \\
                  A_d e^{-i\phi} & -3B_d e^{i\theta} & -3B_d \\
                  0 & -3B_d & C_d \end{array}  \right )
\label{ye}
\end{equation}

The low energy effective theory below GUT thresholds is assumed to be the
Minimal Supersymmetric Standard Model (MSSM).  After the renormalization
effects are taken into account, the model agrees with the data on fermion
mixing for a broad range of the $t$-quark mass, $m_t=150...190 \ GeV$.  In
these fits the phase $\theta$ is relatively
unimportant and can be taken to be zero.

The values of the parameters $|A_q|$ and $C_q$ ($q=u,d$), in (\ref{yu})
and (\ref{yd}) are simply related to the masses of quarks.   On the contrary,
the phase $\phi$, which must have a nonzero value in order for the model to
agree with experiment,  depends on both the masses of quarks and their mixing
parameters.  It is a common procedure to diagonalize the matrices $M_u$ and
$M_d$ numerically (or by means of the Taylor series expansion in terms of the
quark mass ratios) and compare the corresponding CKM matrix to the data.
On the other hand, the method of invariants discussed above offers a simpler
and more elegant solution: the phase $\phi$ can be expressed
analytically in terms
of the measurable quantities only.  Because the mass matrices in (\ref{yu})
and (\ref{yd}) are hermitian for $\theta=0$, one can use a $\tilde{K}$-type
invariant. On one hand, the allowed range for the value of
$\tilde{K}_{11}(1,1)$ is known in terms of the measurable quantities from
(\ref{tilde_kdm}).  On the other hand,

\begin{equation}
\begin{array}{l}
\tilde{K}_{11}(1,1) = det(M_u+M_d) =  \\
=|A_u +A_d | \ (C_u+C_d) =  \\
= (2 | A_u A_d | cos ( \phi ) + |A_u|^2 + |A_d|^2 ) (C_u+C_d)
\end{array}
\label{k_appl}
\end{equation}

And therefore

\begin{equation}
cos(\phi) \ = \frac{1}{2 \: |A_u| \: |A_d|} \ (\tilde{K}_{11}(1,1)/(C_u+C_d)
\ - \ |A_u|^2 \ - \ |A_d|^2 )
\label{phi}
\end{equation}

The right-hand side of (\ref{phi}) is known in terms of the quark masses and
the CKM mixing parameters.  Therefore, one can use the relation (\ref{phi})
to evaluate the phase $\phi$ without having to explicitly diagonalize the mass
matrices.

For this model and others, it is important to have a general methodology to
determine (1) how many unremovable phases there are in quark mass matrices and
(2) which elements of these matrices can be rephased to be real.  We have
presented a general method to answer both of these questions \cite{ksquark}.
The analogous questions for lepton mass matrices are more complicated, but we
have also given a general answer to them \cite{kslepton}.

In summary, we have introduced and studied some new invariants of
the quark mass matrices which are model- and weak basis-independent.  The
identities (\ref{kdm}) and (\ref{ieq}), as well as (\ref{tilde_ieq}) and
(\ref{tilde_kdm}), involving these invariants,
provide important constraints on the possible
forms of quark mass matrices $M_{u}$ and $M_d$
since they directly relate the elements of these matrices to
the measurable parameters $|V_{ij}|^2$ and quark masses, and thereby enable one
to avoid the explicit calculation of the eigenvectors of $H_{u}$ and $H_{d}$.

\section{A method to generate families of viable quark mass matrices}

  In this section we would like to present
some useful mathematical tools, first proposed in \cite{2_tools}, for the
study of quark mass matrices and the connection with quark mixing.
Our method was recently applied in \cite{2_an}
to generate the families of acceptable solutions for the fermion mass
matrices.

The idea of the method is to utilize the well-known experimental fact that the
CKM matrix is close to the $ 3 \times 3 $ identity matrix: $ | V_{ij} |
\approx \delta_{ij} $.  We make use of this fact by writing $V$ as

\begin{equation}
V=e^{i \alpha H} \label{exp}
\end{equation}
where $ H $ is some hermitian matrix and $ \alpha $ is a real number.
One may choose $ H $ to have its dominant (largest, in absolute value)
eigenvalue to be $ 1 $.
Then for $ \alpha $ consistent with the data \cite{3_data}, we find that
$ | \alpha | \approx 0.3 $.  We can now expand $V$ in the powers of $\alpha$:
\begin{equation}
V={\bf 1}+ i \alpha H- \frac{1}{2} \alpha^{2} H^{2} +...
+\frac{1}{n!}(i\alpha H)^{n}+...
\end{equation}

It was shown in Ref. \cite{2_tools} that for any practical purposes
it is sufficient to consider the first and the second order in $\alpha$
to match the precision to which the relevant quantities (CKM parameters,
quark masses, etc.) are known.

The matrix $H$, corresponding to a given matrix $V$ with distinct eigenvalues
$v_i$, can be easily computed using the Sylvester's theorem:

\begin{equation}
i \alpha H=\sum_{k=1}^{3} ln(v_{k})
\frac{\prod_{i \neq k}(V-v_{i} \times {\bf 1})}
{\prod_{i \neq k} (v_{k}-v_{i})}    \label{Sylvester}
\end{equation}

Then the usual system of equations:

\begin{eqnarray}
\left \{ \begin{array}{l}
        U_{u,L} M_{u} U_{u,R}^{\dag}=diag(m_{u},m_{c},m_{t}) \equiv D_{u}  \\
        U_{d,L} M_{d} U_{d,R}^{\dag}=diag(m_{d},m_{s},m_{b}) \equiv D_{d}  \\
        U_{u,L} U_{d,L}^{\dag}=V
         \end{array}    \right.
\end{eqnarray}
is satisfied for the family of solutions:
\begin{eqnarray}
\left \{ \begin{array}{l}
M_{u} =U_{u}^{\dag} D_{u} U_{u} =        \\
D_{u}+i \alpha x [D_{u},H]-\frac{1}{2} \alpha^{2} x^{2} [[D_{u},H],H]+... \\
 \\
M_{d} =U_{d}^{\dag} D_{d} U_{d}=       \\
D_{d}+i \alpha (x-1) [D_{d};H]-
\frac{1}{2} \alpha^{2} (x-1)^{2} [[D_{d},H],H]+...
         \end{array}    \right.
\end{eqnarray}
depending on some arbitrary parameter $x$. (It is assumed that $|x|$, as well
as $|1-x|$, is sufficiently small to preserve the convergence of the series.)

To summarize, in this presentation we have discussed two different approaches
to analyzing quark mixing.  First, we introduced some new invariants
of quark mixing and showed the usefulness of our method of invariants on a
simple example.  Then we have also discussed some new mathematical tools which
allow one to generate the mass matrices consistent with the data.

The author would like to thank Professor Robert Shrock for many helpful
discussions and comments.

\end{document}